\documentclass[10pt]{article}

\usepackage{graphicx}
\usepackage{fancyheadings}

\begin{document}

\pagestyle{fancy}
\chead{{\it CMB Component Separation and the Physics of Foregrounds \\ 14-18 July 2008, Pasadena, California}}

\centerline{\sc \Large Anomalous emission from HII regions}
\vspace{.5pc}
\centerline{\sc Clive Dickinson$^{1}$}
\centerline{\it Infrared Processing \& Analysis Center, California Institute of Technology}
\vspace{.5pc}

\normalsize

\abstract{Spinning dust appears to be the best explanation for the anomalous emission that has been observed at $\sim 10-60$~GHz. One of the best examples of spinning dust comes from a HII region in the Perseus molecular cloud. Observations of other HII regions also show tentative evidence for excess emission at frequencies $\sim 30$~GHz, although at lower emissivity levels. A new detection of excess emission at 31~GHz in the HII region RCW175 has been made. The most plausible explanation again comes from spinning dust. HII regions are a good place to look for spinning dust as long as accurate radio data spanning the $\sim 5-100$~GHz range is available. }

\vspace{0.75pc}
{\bf \large 1. Introduction}
\vspace{0.5pc}

Anomalous emission appears to be ``real'' i.e. there is an additional component of diffuse Galactic microwave emission that is not traditional synchrotron, free-free or thermal dust emissions (e.g. Davies et al. 2006). The anomalous emission emits strongly in the frequency range $\sim 10-60$~GHz and is strongly correlated with FIR dust. A number of physical mechanisms could be responsible for the anomalous emission including hard synchrotron, hot free-free emission, cold dust, magneto-dipole radiation, fullerenes, to name a few. However, so far, the best explanation for the physical mechanism responsible for the anomalous emission appears to be electro-dipole radiation from ultra-rapidly spinning dust grains, often referred to as ``spinning dust'' (Draine \& Lazarian 1998a,b).

\vspace{0.75pc}
{\bf \large 2. Why HII regions?}
\vspace{0.5pc}

HII regions may be a good place to look for anomalous emission. Most importantly, they are typically associated with large column densities of dust, and therefore emit a detectable spinning dust signal. Even though they emit strong free-free emission from the warm ionized gas ($T_{e}\approx 10000$~K), this component can be quantified either by its well-defined spectrum (a power-law with a spectral index $\beta_{ff}=-0.15\pm 0.05$) or alternatively, from recombination lines (e.g. H$\alpha$, RRLs) (Dickinson et al. 2003). Draine \& Lazarian (1998b) considered six typical  environments (CNM,WNM,WIM etc.) and they found that the the emissivity of spinning dust is not strongly dependent on the details of the environment, as shown in Fig. 1. There is of course a contrary argument regarding small grain destruction in the environments around hot starts that form HII regions, particularly the PAHs that could be spinning at fast rates. However, this is usually only in the central region of a compact HII region and the dense dust population in the surrounding material (which is usually not resolved by telescopes with beams $>>$ arcmin) is typically not strongly affected. It is interesting to note that the dust temperature is typically 30-50~K, compared to the $\approx 18$~K dust found in the diffuse ISM.

\begin{figure*}[!h]
\centering \includegraphics[width=0.7\textwidth, angle=0]{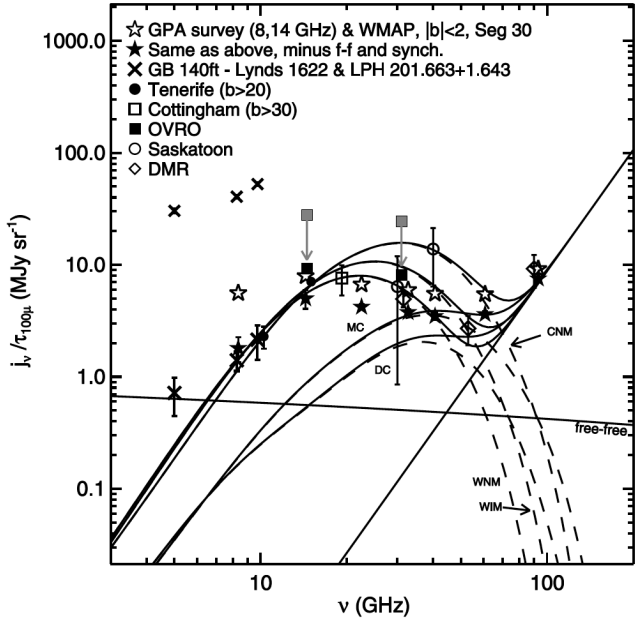}
\caption{Model dust emissivities from Draine \& Lazarian (1998) with previous observational data, reproduced from Finkbeiner et al. (2004). Note how the spinning dust models for different environments (CNM, WNM, WIM etc.) peak in the range $\sim20-40$~GHz and typically vary by only a factor of a few in emissivity.}
\label{fig:spdust_specs}
\end{figure*}

\vspace{0.75pc}
{\bf \large 3. Detections from ``compact'' Galactic objects}
\vspace{0.5pc}

The best two examples of spinning dust are the HII region G159.6-18.5 in the Perseus molecular cloud (Watson et al. 2005) and the dark cloud LDN1622 (Finkbeiner et al, 2002,2004; Casassus et al. 2006). Both of these show spectra that are consistent with canonical spinning dust models. Furthermore, LDN1622 shows a better morphological correlation with the shorter wavelengths of IRAS (12/25$~\mu$m), as expected from the smaller dust grains. 

Of the two spinning dust candidates of Finkbeiner et al. (2002), the strongest candidate was initially LPH[96]201.663+1.643. Follow-up observations with the Cosmic Background Imager (CBI) showed that the flux density at 31~GHz was completely consistent with optically thin free-free emission. The original Finkbeiner et al. (2002) data at 5-10~GHz and the CBI 31~GHz data point can still be fit by an anomalous component with a very narrow spectrum (N. Ysard, priv. comm.). However, more recent GBT 100~m observations (D. Finkbeiner, priv. comm.) could not reproduce the original Green Bank $5-10$~GHz data, suggesting that the earlier data were spurious.

Dickinson et al. (2007) observed 6 bright southern HII regions with the CBI at 31~GHz to search for anomalous emission. Using low frequency data from the literature, a simple free-free power-law was fitted to the spectrum of each object. The 31~GHz flux density appeared to be slightly above the simple free-free extrapolation. The average dust emissivity, relative to the $100~\mu$m map, was found to be $3.3\pm1.7~\mu$K/(MJy/sr). This is significantly lower than values observed at high latitudes and other anomalous regions (Table 1). On the other hand, Scaife et al. (2007) observed a sample of 16 HII regions with the Arc Minute Imager (AMI) at $\approx 15$~GHz, and found no significant evidence for excess emission with many regions emitting at $<1~\mu$K/(MJy/sr).

\begin{table}
\caption{Comparison of $100~\mu$m dust emissivities for HII~regions and cooler dust clouds, from data at or near 30~GHz. Emissivities, in units $\mu$K~(MJy/sr)$^{-1}$, have been normalised to 31~GHz.}
\vspace{0.5pc}
\small
\begin{center}
\begin{tabular}{lcl} \hline
Source   &Dust emissivity         &Reference \\
         &$\mu$K~(MJy/sr)$^{-1}$  &       \\
\hline
{\bf HII~regions}    &             &       \\
6 HII~regions (mean) &$3.3\pm1.7$  &Dickinson et al. (2007)       \\
\vspace{1mm}
LPH96                 &$5.8 \pm 2.3$&Dickinson et al. (2006) \\

{\bf Cool dust clouds} &             &       \\

15 regions {\it WMAP} &$11.2\pm1.5$ &Davies et al. (2006) \\

All-sky {\it WMAP}    &$10.9\pm1.1$ &Davies et al. (2006) \\

LDN1622               &$24.1\pm0.7$ &Casassus et al. (2006) \\

$G159.618.5$          &$17.8\pm0.3$ &Watson et al. (2005) \\
\hline
\end{tabular}
\label{tab:emissivities}
\end{center}
\end{table}

\normalsize

\vspace{0.75pc}
{\bf \large 4. New detection of spinning dust in RCW175}
\vspace{0.5pc}

Preliminary results from the Very Small Array (VSA) Galactic plane survey at 33~GHz (Todorovic et al, in prep.) indicated that the RCW175 HII region was significantly brighter than expected from lower frequency surveys. It was re-observed with higher resolution with the CBI at 31~GHz and a consistent flux density was measured (Dickinson et al. 2008). The integrated flux density spectrum of RCW175 is shown in Fig. 2. Best-fitting models for free-free and thermal dust emission are plotted along with a typical spinning dust model. The upper limit at 94~GHz from WMAP data is very important for the interpretation of the apparent excess since it effectively rules out significant contributed from thermal dust or optically thick free-free emission, perhaps from an ultracompact HII region. The estimated column density from IRAS $100~\mu$m maps gives a spinning dust emissivity that is consistent with the Draine \& Lazarian (1998) models.

\begin{figure*}[!h]
\centering \includegraphics[width=0.7\textwidth, angle=0]{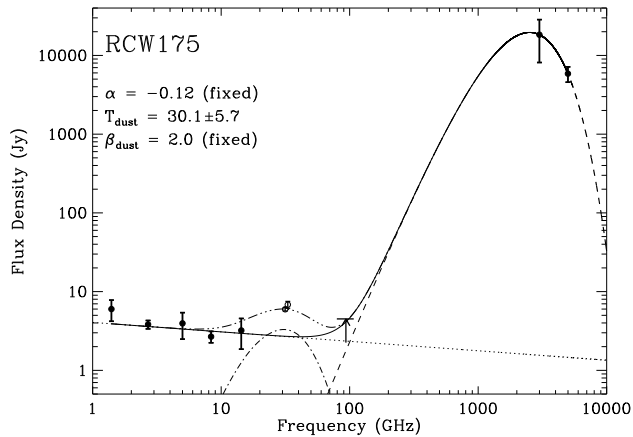}
\caption{Integrated flux density spectrum for RCW175 (Dickinson et al. 2008). Filled circles represent data fitted by a power-law with fixed spectral index (dotted line) and 
a modified black-body with fixed emissivity (dashed line). The Draine \& Lazarian (1998b) CNM spinning dust spectrum (dot
-dashed line) has been fitted to the 31/33~GHz data.  }
\label{fig:spec}
\end{figure*}

\vspace{0.75pc}
{\bf \large 5. Conclusion}
\vspace{0.5pc}

Anomalous emission has been observed at both high latitudes  and from discrete Galactic objects. The best explanation is that it is due to spinning dust grains. One of the best examples of spinning dust is the HII region G159.6-18.5 in the Perseus molecular cloud. There is also tentative evidence for excess emission from other HII regions. In particular, RCW175 has now been confirmed to have excess emission at $\sim 30$~GHz with spinning dust being a plausible origin. HII regions are good places to search for spinning dust emission, but accurately calibrated data covering the frequency range $\sim 5-100$~GHz is required.

\vspace{0.75pc}
{\bf References}
\vspace{0.5pc}
\footnotesize \\
Casassus, S., et al., 2006, ApJ, 639, 951 \\
Davies, R.D.,  et al., 2006, MNRAS, 370, 1125 \\
Dickinson, C., Davies, R.D., Davis, R.J., 2003, MNRAS, 341, 369 \\
Dickinson, C., et al., 2006, ApJ, 643, L111 \\
Dickinson, C., et al., 2007, MNRAS, 379, 297 \\
Dickinson, C., et al., 2008, ApJL, submitted (arXiv:0807.3985) \\
Draine, B.T., Lazarian, A., 1998a, ApJ, 494, L19 \\
Draine, B.T., Lazarian, A., 1998b, ApJ, 508, 157 \\
Finkbeiner, D., et al., 2002, ApJ, 566, 898 \\
Finkbeiner, D., Langston, G.I., Minter, A.H., 2004, ApJ, 617, 350 \\
Scaife, A., et al., 2007, MNRAS, 385, 809 \\
Watson, R.A., et al., 2005, ApJ, 624, L89

\end{document}